\documentclass[prd,article,nofootinbib,twocolumn,preprintnumbers,superscriptaddress]{revtex4}

\usepackage{rotating}
\usepackage{amsmath,amstext,amsbsy,amscd,bbm,epsfig,lscape}
\usepackage{graphicx,amsfonts,amssymb,color,comment,float,times}

\usepackage[utf8]{inputenc}

\definecolor{MyGrey}{rgb}{0,0,0} 
\definecolor{MyDarkBlue}{rgb}{0.,0.,1} 
\definecolor{MyLightBlue}{rgb}{0.22,0.51,0.9}

\usepackage[colorlinks=true,linkcolor=MyDarkBlue,citecolor=MyDarkBlue,urlcolor=MyLightBlue,bookmarksnumbered=true,bookmarksopen]{hyperref}

\usepackage{mathrsfs,amssymb,slashed}  
\usepackage{cancel}
\usepackage[normalem]{ulem}

\newcommand{\be}{\begin{equation}}
\newcommand{\ee}{\end{equation}}
\newcommand{\bea}{\begin{eqnarray}}
\newcommand{\eea}{\end{eqnarray}}

\begin{document}

\preprint{FERMILAB-PUB-17-287-T, IFT-UAM/CSIC-17-071}

\title{Double Bangs from New Physics in IceCube}

\author{Pilar Coloma}
\email{pcoloma@fnal.gov}
\affiliation{Theory Department, Fermi National Accelerator Laboratory, P.O. Box 500, Batavia, IL 60510, USA}
\author{Pedro A.~N. Machado}
\email{pmachado@fnal.gov}
\affiliation{Theory Department, Fermi National Accelerator Laboratory, P.O. Box 500, Batavia, IL 60510, USA}
\author{Ivan Martinez-Soler}
\email{ivanj.m@csic.es}
\affiliation{Instituto de Fisica Teorica UAM-CSIC, Calle Nicolas Cabrera 13-15, Universidad Autonoma
de Madrid, Cantoblanco, E-28049 Madrid, Spain}
\author{Ian M. Shoemaker}
\email{ian.shoemaker@usd.edu}
\affiliation{Department of Physics, University of South Dakota, Vermillion, SD 57069, USA}

\date{\today}
\begin{abstract}
A variety of new physics models allows for neutrinos to up-scatter into heavier states. If the incident neutrino is energetic enough, the heavy neutrino may travel some distance before decaying. In this work, we consider the atmospheric neutrino flux as a source of such events. At IceCube, this would lead to a ``double-bang'' (DB) event topology, similar to what is predicted to occur for tau neutrinos at ultra-high energies. The DB event topology has an extremely low background rate from coincident atmospheric cascades, making this a distinctive signature of new physics. Our results indicate that IceCube should already be able to derive new competitive constraints on models with GeV-scale sterile neutrinos using existing data. 
\end{abstract}
\preprint{}


\maketitle

\textit{Introduction.} 
Although neutrino physics has rapidly moved into the precision era, a number of fundamental questions remain unanswered. Perhaps the most important among these is the mechanism responsible for neutrino masses. In the most na\"\i ve extension of the Standard Model (SM), neutrino masses and mixing can be successfully generated by adding at least two right-handed neutrinos ($N_R$), with small Yukawa couplings $Y_\nu$ to the left-handed lepton doublets $L_L$ and the Higgs boson $\phi$. In this framework, Dirac neutrino masses are generated after Electro-Weak (EW) symmetry breaking, as for the rest of the SM fermions. As singlets of the SM the right-handed neutrinos may also have a Majorana mass term, since it is allowed by gauge symmetry. In this case, the neutrino mass Lagrangian reads
\[
\mathcal{L}_{mass}^{\nu} \supset Y_{\nu} \overline{L}_L \tilde\phi N_R + \frac{1}{2}M_R \overline{N_R^c}N_R \, + \textrm{h.c.} \, ,
\]
where $\tilde\phi \equiv i\sigma_2 \phi^*$, $ N_R^c \equiv C \bar N_R^T$ is the charge conjugate of $N_R$ and we have omitted flavor and mass indices. This is the well-known Type I Seesaw Lagrangian\cite{Minkowski:1977sc,GellMann:1980vs,Mohapatra:1979ia}. Traditionally, the Type I Seesaw assumed a very high Majorana mass scale $M_R$. For $M_R\gg v$ the light neutrino masses are proportional to $m_\nu \propto Y_\nu^\dagger M^{-1}_R Y_\nu v^2$, where $v$ is the Higgs vacuum expectation value, while the right-handed neutrino masses would be approximately $m_N \simeq M_R + \mathcal{O}(m_\nu)$. In this framework the SM neutrino masses are naturally suppressed by the new physics scale and can be much smaller than the charged fermion masses without the need for tiny Yukawa couplings. However, such heavy neutrinos are too heavy to be produced in colliders, and the inclusion of very massive Majorana neutrinos would considerably worsen the hierarchy problem for the Higgs mass~\cite{Vissani:1997ys}.

Models with lower values of $m_N$ can lead to a more interesting phenomenology, testable at low-energy experiments, and possibly even solve some of the other problems of the SM. For example, keV neutrinos offer a very good dark matter candidate~\cite{Kusenko:2009up}, while Majorana neutrinos with masses $m_N\sim\mathcal{O}(1-100)$~GeV can successfully generate the matter-antimatter asymmetry of the Universe~\cite{Asaka:2005an,Asaka:2005pn,Hernandez:2016kel,Hernandez:2015wna}. 
While right-handed neutrinos with masses above the EW scale are subject to very tight bounds from EW observables and charged lepton flavor violating experiments~\cite{Antusch:2014woa,Fernandez-Martinez:2016lgt}, these constraints fade away for lower masses. Indeed, for right-handed neutrinos in the (keV - GeV) range the strongest constraints come from precision measurements of meson decays, muon decays and other EW transitions, see \textit{e.g.} Ref.~\cite{Atre:2009rg} for a review. 

\begin{figure}[t!]
\begin{center}
\includegraphics[width=\columnwidth]{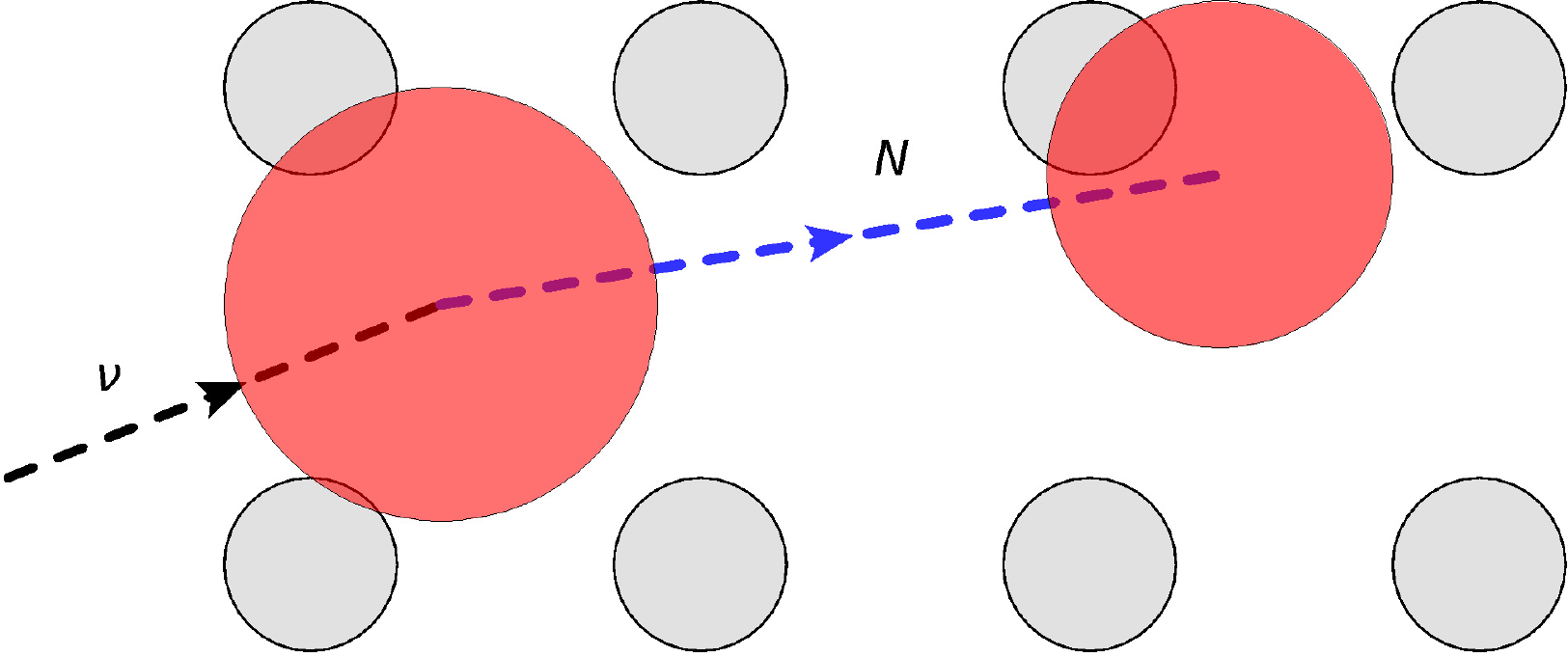}
\caption{Schematic illustration of a DB event in IceCube. An incoming active neutrino $\nu$ up-scatters into a heavy neutrino $N$, which then propagates and decays into SM particles. The small circles represent the DOMs while the large circles indicate the positions where energy was deposited. \label{fig:scheme} }
\end{center}
\end{figure}

In this letter we point out that IceCube and DeepCore can be used to test models with GeV neutrinos directly. To this end, we consider events with a ``double-bang'' (DB) topology. A schematic illustration of the event topology can be seen in Fig.~\ref{fig:scheme}. In the first interaction, an atmospheric neutrino would up-scatter off a nucleus into a heavier state. This generally leaves a visible shower (or cascade) in the detector coming from the hadronic part of the vertex. After traveling a macroscopic distance inside the instrumented ice, the heavy neutrino would decay back to SM particles. The decay will produce a second cascade if the final state involves charged particles or photons which can be detected by IceCube's digital optical modules (DOMs). Thus, the final DB topology would be two cascades (or ``bangs'') visibly separated. This topology is predicted to occur in the SM from the production of a $\tau$ lepton in $\nu_{\tau}$ charged-current (CC) scattering at PeV energies~\cite{Learned:1994wg}, and has already been searched for by the collaboration~\cite{Aartsen:2015dlt}.  In our case, however, the heavy neutrinos will be produced from the atmospheric neutrino flux and thus produce much lower energy DBs. 


To illustrate some of the new physics scenarios giving rise to low-energy DB events we consider two basic scenarios depending on the main production/decay mode of the heavy state: (1) through mixing with the light neutrinos, and (2) through a transition magnetic moment involving the light neutrinos.

{\it Heavy neutrino production via mixing.} The measurement of the invisible decay width of the $Z$ implies that, if additional neutrinos below the EW scale are present, they cannot couple directly to the $Z$ (\emph{i.e.,} they should be ``sterile''). For simplicity, let us focus on a scenario where there is sizable mixing with only one heavy neutrino while the others are effectively decoupled. We may write the flavor states $\nu_\alpha$ as a superposition of the mass eigenstates as
\be \nu_{\alpha L} = \sum_{i=1}^{3} U_{\alpha i} \nu_{i L} + U_{\alpha 4} N_{4 R}^{c} \, ,
\ee
where $U$ is the $4\times 4$ unitary mixing matrix that changes between the mass and the flavor bases. For a sterile neutrino with a mass $m_N \sim \mathcal{O}(0.1 - 10)~{\rm GeV}$, its mixing with $\nu_{e,\mu}$ is severely constrained as $|U_{\alpha 4}|^2 \lesssim 10^{-5} - 10^{-8}$ ($\alpha = e,\mu$)~\cite{Atre:2009rg}. Conversely, the mixing with $\nu_\tau$ is much more difficult to probe, given the technical challenges of producing and detecting tau neutrinos. For $m_N \sim \mathcal{O}(0.1-10)~\textrm{GeV}$ the most stringent bounds are derived from the DELPHI~\cite{Abreu:1996pa} and CHARM~\cite{Orloff:2002de} experiments. However, a mixing as large as $|U_{\tau 4}|^2 \sim 10^{-2}$ is still allowed for masses around $m_N \sim \mathcal{O}(400)~\textrm{MeV}$~\cite{Atre:2009rg}.

At IceCube, the atmospheric neutrino flux can be used to constrain the values of $U_{\alpha 4}$ directly. Atmospheric neutrinos are produced as a result of the cosmic rays impacting the atmosphere. At the production point, this flux is primarily composed of $\nu_\mu$ and $\nu_e$. However, for neutrinos crossing the Earth a large fraction of the initial $\nu_\mu$ flux will have oscillated into $\nu_\tau$ by the time the neutrinos reach the detector. Therefore, here we focus on probing the mixing with $\nu_\tau$ since this one is much harder to constrain by other means. 


To this end, we propose to conduct a search for low-energy DB events. In each event the first cascade is produced from a neutral-current (NC) interaction with a nucleon $n$, as $\nu \;  n \to N \; n$. This process is mediated by a $Z$ boson and takes place via mixing between the light and heavy states. Neglecting corrections due to the mass of the heavy neutrino, the up-scattering cross section goes as $\sigma_{\nu_\tau N} \simeq \sigma_\nu^{NC}\times | U_{\tau 4} |^2 $ where $\sigma_\nu^{NC}$ is the NC neutrino-nucleon cross section in the SM. Unless the process is quasi-elastic, it will generally lead to a hadronic shower in the detector. Here we compute the neutrino-nucleon deep-inelastic scattering (DIS) cross section using the parton model, imposing a lower cut on the hadronic shower of 5~GeV so it is observable~\cite{Aartsen:2017nmd}. Once the heavy state has been produced, its decay is controlled by kinematics and the SM interactions inherited from the mixing with the active neutrinos. The partial decay widths of a heavy neutrino can be found in Refs.~\cite{Gorbunov:2007ak,Atre:2009rg} and were recomputed here. The decay channels include two-body decays into a charged lepton (active neutrino) and a charged (neutral) meson, and three body decays into charged leptons and light neutrinos. The deposited energy in the second shower is also required to be above 5~GeV. It should be noted that if the $N$ decays into three light neutrinos the second shower will be invisible: those events do not contribute to our signal. As an example, for $m_N = 1$~GeV and $|U_{\tau 4}|^2 = 10^{-3}$, the boosted decay length (for an energy of 10~GeV) is $L_{lab}\sim 20$m.  

The number of DB events from $\nu_\tau$ mixing with a heavy neutrino, for two cascades taking place within a distance $L$, is proportional to
\be
\!\int \! dE_\nu dc_{\theta} \mathcal{B} \frac{d \phi_{\nu_\mu}}{dE_\nu dc_{\theta}}~P_{\mu\tau}(c_\theta, E_\nu)~\frac{d \sigma_{\nu_{\tau}N}}{dE_\nu}~P_{d}(L)~V (L,c_\theta),
\label{eq:events}
\ee
where $E_\nu$ is the incident neutrino energy and $c_\theta \equiv \cos\theta$ is the cosine of its zenith angle. The atmospheric $\nu_\mu$ flux~\cite{Honda:2015fha} is given by $\phi_{\nu_\mu}$ while $P_{\mu\tau}$ is the oscillation probability in the $\nu_\mu \to \nu_\tau$ channel, which depends on the length of the baseline traveled (inferred from the zenith angle) and the energy. Here, $P_d(L) = e^{-L /L_{lab}}/L_{lab}$ is the probability for the heavy state to decay after traveling a distance $L$, while $ \mathcal{B}$ is its branching ratio into visible final states (\textit{i.e.}, excluding the decay into three light neutrinos). Antineutrino events will give a similar contribution to the total number of events, replacing $ \phi_{\nu_\mu}$, $\sigma_{\nu_\tau N}$ and $P_{\mu\tau}$  in Eq.~\eqref{eq:events} by their analogous expressions for antineutrinos. 

In Eq.~\eqref{eq:events} we have omitted a normalization constant which depends on the number of target nuclei and the data taking period, but we explicitly include an effective volume $V (L,c_\theta)$. In this work, this was computed using Monte Carlo integration. First, for triggering purposes we require that at least three (four) DOMs detect the first shower simultaneously, if it takes place inside (outside) DeepCore~\cite{Aartsen:2016nxy}. Once the trigger goes off, all the information in the detector is recorded, and we thus assume that the second shower is always observed as long as it is close enough to a DOM.  
Eventually, the energy of a cascade determines the distance from which it can be detected by a DOM: the longer the distance, the more energetic the cascade should be so the light can reach the DOM without being absorbed by the ice first. Here we assume that a cascade is seen by a DOM if it takes place within a distance of 36~m~\cite{Aartsen:2016nxy}. This is conservative, since showers with energies much above 5~GeV will typically reach a DOM from longer distances. 
Finally, a minimum separation is required between the two showers so they can be resolved. This ultimately depends on the time resolution of the DOMs. Following Ref.~\cite{Aartsen:2015dlt}, we require a minimum distance between the two showers of 20~m.

The dominant source of background for DB events is given by two coincident cascades taking place within the same time window $\Delta t$. The rate can be estimated as~\cite{Ge:2017poy} $N_{bkg} \simeq  C_{DB}^{2}~ (\Delta t/T)^{2}$, where $C_{DB}^{2} = N_{casc} (N_{casc}-1)/2$ comes from the number of possible combination of pairs, and $N_{casc}$ is the number of cascade events within a time period $T$. The number of cascades in the DeepCore volume, with a deposited energy between 5.6~GeV and 100~GeV, is $ N_{casc}\simeq 2 \times 10^{4}~{\rm yr}^{-1}$~\cite{LakeLouiseTalk}. These include CC events with electrons, taus or low-energy muons in the final state (which do not leave long identificable tracks), as well as NC events. Thus, for $\Delta t = 500 $~s, we get approximately $N_{bkg} \simeq 0.05~{\rm yr}^{-1}$. However this estimate is probably conservative, as a shorter $\Delta t$ may be used for the low energies considered here. 

\begin{figure}[t!]
\begin{center}
\includegraphics[width=\columnwidth]{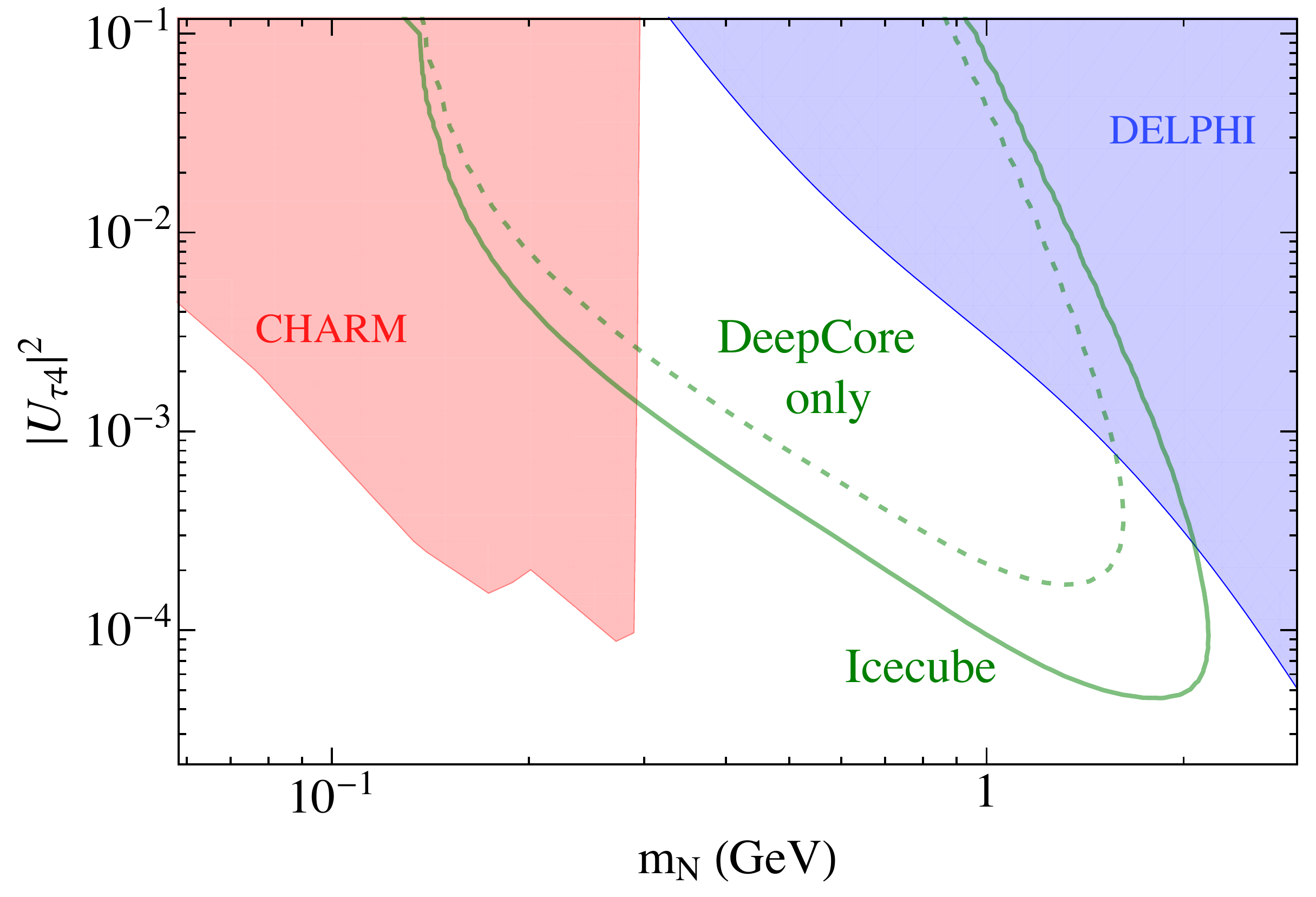}
\caption{Expected potential of IceCube to constrain the mixing between $\nu_\tau$ and a heavy neutrino. In the region enclosed by the solid green contour, more than one DB events are expected during 6 years of data taking at IceCube. The dashed contour shows the most conservative result where only the DeepCore volume is considered. The shaded regions are disfavored by CHARM~\cite{Orloff:2002de} and DELPHI~\cite{Abreu:1996pa} at 90\% and 95\% CL respectively, see Ref.~\cite{Atre:2009rg}. \label{fig:sens-U} }
\end{center}
\end{figure}

In view of the negligible background rate, we proceed to determine the region in parameter space where at least one signal event would be expected in six years of data taking at IceCube. This is shown in Fig.~\ref{fig:sens-U} as a function of the mass and mixing of the heavy neutrino. The solid line shows the results using the full IceCube volume, while for the dashed line only DeepCore was considered. Our results indicate that IceCube could improve over present bounds between one and two orders of magnitude, and probe values of the mixing as small as $|U_{\tau 4}|^2 \sim 5 \times 10^{-5}$. According to these results, IceCube could test the proposed solution to the flavor anomalies in the $B$ sector proposed in Ref.~\cite{Cvetic:2017gkt}.


{\it Heavy neutrino production via a transition magnetic moment.} Alternatively, the light neutrinos may interact with the heavy state $N$ through a higher-dimensional operator. As an example, we consider a neutrino transition magnetic moment (NTMM) $\mu_\textrm{tr}$:
\be \mathcal{L}_\nu \supset -\mu_\textrm{tr} ~\overline{\nu}_{\alpha L} \sigma_{\rho \sigma} N_{4R} F^{ \rho \sigma},
\label{eq:operator}
\ee
where $F^{\rho \sigma}$ is the electromagnetic field strength tensor and $\sigma_{\rho \sigma} = \frac{i}{2}[\gamma_\rho, \gamma_\sigma]$. For simplicity, in this scenario we assume negligible mixing with the light neutrinos, so both the production and decay of the heavy neutrino are controlled by the magnetic moment operator. In the rest frame of $N$, its decay width reads $\Gamma(N \to \nu \gamma) = \mu_\textrm{tr}^{2} M^{3}/(16 \pi)$. For $m_N=100$~MeV, $\mu_{\rm tr}=10^{-8}\mu_B$ (where $\mu_B$ is the Bohr magneton), and a typical energy of 10~GeV this gives a decay length in the lab frame $L_{lab}\sim 14$~m.

Neutrinos with a NTMM could scatter off both electrons and nuclei in the IceCube detector. However, for the range of energies and masses considered in this work, the largest effect comes from scattering on nuclei. In the DIS regime, the cross section for the scattering $\nu \; n \to N \; n$ via the operator in Eq.~\eqref{eq:operator} reads
\begin{equation}
\label{eq:TNMM-xsec}
  \frac{d^2\sigma_{\nu n \to  N n}}{dxdy}\simeq 16\pi\alpha \mu_{\rm tr}^2 \left(\frac{1-y}{y}\right) \sum_i e_i^2 f_i(x),
\end{equation}
where $\alpha$ is the fine structure constant, $f_i(x)$ is the parton distribution function for the parton $i$, $x$ is the parton momentum fraction and $e_i^2$ is its electric charge. Here, $y\equiv 1-E_N / E_\nu = E_r/E_\nu$, where $E_N$ is the energy of the outgoing heavy neutrino and $E_r$ is the deposited energy. In Eq.~\eqref{eq:TNMM-xsec} we have ignored the impact of the heavy neutrino mass in the cross section, which will be negligible in the region of interest. However, energy and momentum conservation do require the following inequality to be satisfied:
\begin{align}
E_r^2-W^2 - &\big[ m_N^2-W^2-2x E_\nu m_n   \label{eq:maxmass-2}  \\
  & - x^2m_n^2 +2E_r(x m_n+E_\nu)\big]^2/4E_\nu^2>0 \, , \nonumber
\end{align}  
where $W^2$ is the invariant mass squared of the outgoing hadronic system and $m_n$ is the nucleon mass. Using the DIS cross section in Eq.~\eqref{eq:TNMM-xsec} and imposing the restriction in Eq.~\eqref{eq:maxmass-2} we can estimate the number of DB events in IceCube using a similar expression to Eq.~\eqref{eq:events}. A 5~GeV lower cut is also imposed on the deposited energy for each shower. Assuming that the decay only takes place via NTMM, the branching ratio to visible final states in this scenario is $\mathcal{B} = 1$. 

\begin{figure*}[t!]  
\centerline{
  \includegraphics[width=\columnwidth]{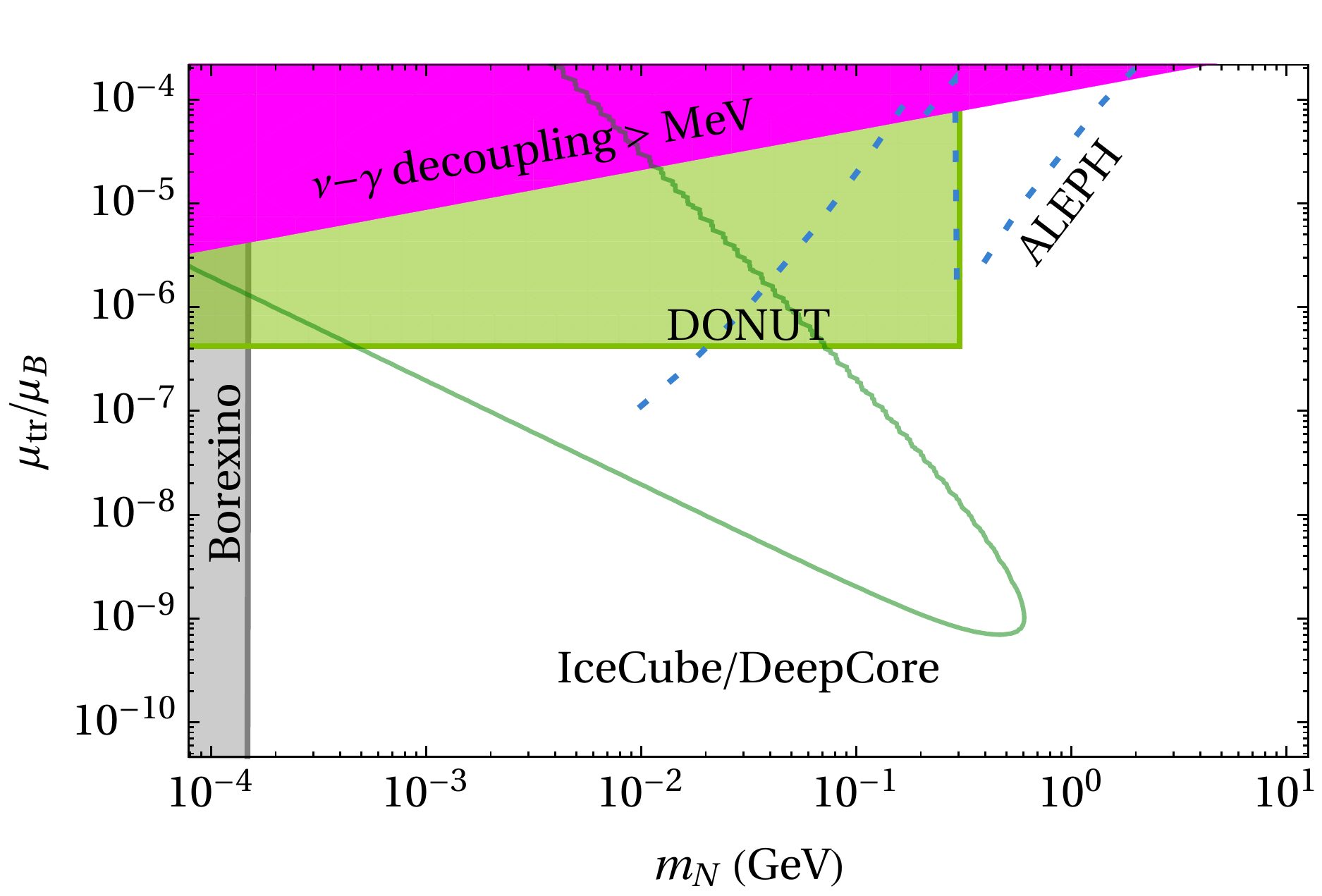}
  \includegraphics[width=\columnwidth]{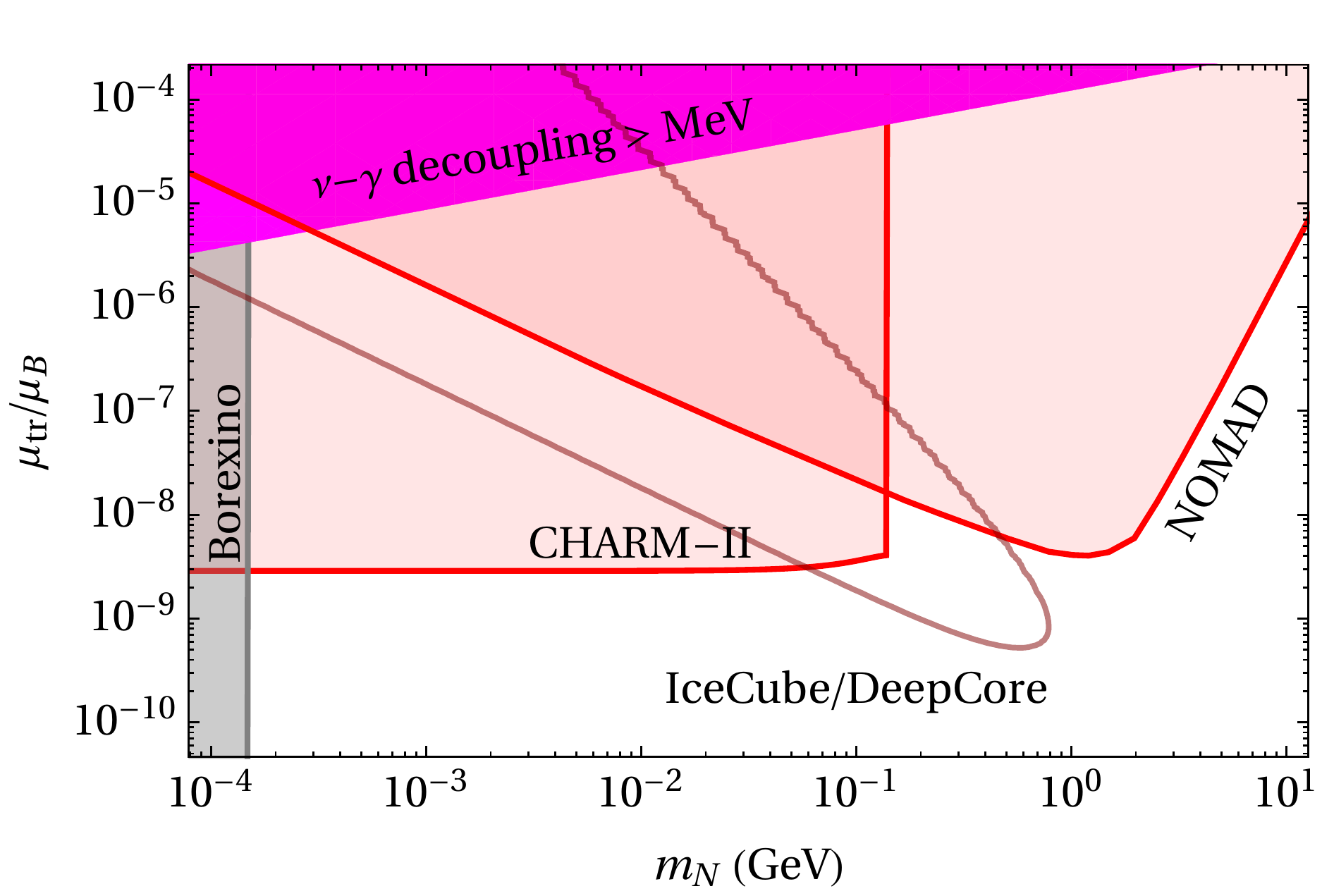}}
  \caption{Expected potential to constrain magnetic moments leading to the transitions $\nu_\tau - N$ (left panel) and $\nu_\mu - N$ (right panel) at IceCube. In the region enclosed by the solid contours, at least one DB event would be expected at IceCube, for a data taking period of six years. The shaded regions are disfavored by previous experiments, see text for details. \label{fig:sens-mu}}
\end{figure*}

Before presenting our results, let us first discuss the current constraints on NTMM. Previous measurements of the neutrino--electron elastic scattering cross section can be translated into a bound on NTMM. The cross section for neutrino--electron scattering via a NTMM reads
\begin{equation}
\begin{split}
  \frac{d\sigma_{\nu e\to N e}}{dE_r} = \mu_{\rm tr}^2 \alpha & \left[  \frac{1}{E_r} - 
  \frac{m_N^2}{2E_\nu E_r m_e}\left(1-\frac{E_r}{2E_\nu} 
  +\frac{m_e}{2E_\nu}\right) -
   \right. \\ & \left.  
 -\frac{1}{E_\nu} +\frac{m_N^4(E_r-m_e)}{8E_\nu^2E_r^2m_e^2}\right],
\end{split}
\end{equation}
where $m_e$ is the electron mass. Moreover, for given $E_\nu$ and $E_r$, the maximum $m_N$ allowed by kinematics is 
\begin{equation}
  m_{N\!,{\rm max}}^2=2\left[ E_\nu\sqrt{E_r(E_r+2m_e)}-E_r(E_\nu+m_e)\right].
\label{eq:maxmass}
\end{equation}
Table~\ref{tab:experiments} summarizes the past experimental results that would be sensitive to a NTMM through its impact on neutrino--electron scattering. For DONUT we use their constraint on $\nu_\tau$ magnetic moment, $\mu_\tau<3.9\times 10^{-7}\mu_B$~\cite{Schwienhorst:2001sj}. For the rest of the experiments listed, we have derived an approximate limit requiring the NTMM cross section to be below the reported precision on the measurement of the neutrino--electron cross section\footnote{Neutrino--nucleus scattering would in principle be sensitive to NTMM as well. However, the approximate bounds derived are not as competitive. For example, using NuTeV data~\cite{Zeller:2001hh} we find an approximate bound $\mu_{\textrm{tr}} \lesssim 10^{-4}\mu_B$. }, using as ``typical'' values of $E_\nu$ and $E_r$ the ones in Table~\ref{tab:experiments}. For NOMAD~\cite{Altegoer:1997gv} we use the results from Ref.~\cite{Gninenko:1998nn}, in which the rate of Primakoff conversion $\nu_\mu+X\to\nu_s+X(+\gamma)$ (where $X$ is a nucleus) was used to constrain  the transition magnetic moment.

\begin{table}[b!]
  \begin{center}
\renewcommand\arraystretch{1.3}
    \begin{tabular}{|c|c|c|c|c|c|c|} \hline
Measurement  & Flavor &$E_\nu$ &$E_r$ & $m_{N\!,{\rm max}}$ & $\Delta \sigma_{exp}$\\ \hline
Borexino-pp~\cite{Bellini:2014uqa}  & all & 420 keV &230 keV&150~keV & 0.1\\ \hline
Borexino-$^7$B~\cite{Bellini:2014uqa}  & all & 862 keV & 600 keV& 230~keV & 0.1\\ \hline
CHARM-II~\cite{Geiregat:1989sz}  & $\nu_\mu$ & 24 GeV & 5 GeV &  140 MeV & 0.44\\ \hline
DONUT~\cite{Schwienhorst:2001sj}  & $\nu_\tau$ & 100 GeV & 20 GeV & 300 MeV & Ref.~\cite{Schwienhorst:2001sj}\\ \hline
    \end{tabular}
  \end{center}	
  \caption{Set of past experiments that provide competitive constraints on NTMM through measurements of $\nu-e$ scattering. For reference, we list the neutrino flavor and the typical values of $E_\nu$ and $E_r$ used in each experiment, together with the maximum heavy neutrino mass allowed by kinematics (given by Eq. \eqref{eq:maxmass}) and the reported precision on the cross section measurement. 
\label{tab:experiments}  }
\end{table}

The ALEPH constraint on the branching ratio ${\rm BR}(Z\to\nu N\to \nu\nu\gamma)<2.7\times10^{-5}$~\cite{Decamp:1991uy} translates into the bound $|U_{\alpha 4}|^2(\mu_{\rm tr}/\mu_B)^2<1.9\times10^{-16}$~\cite{Gninenko:2010pr}, for $m_N$ below the $Z$ boson mass and $\alpha \equiv e,\mu,\tau$. It can be evaded by requiring a very small mixing with the active neutrinos; however, saturating the bound from direct searches on the mixing $|U_{\tau 4}|^2$ gives the strongest possible constraint from ALEPH data, which is competitive in the mass region $m_N \gtrsim 5-10$~GeV.

Additional bounds on $\mu_{\rm tr}$ can also be derived from cosmology. In the SM, neutrino decoupling takes places at temperatures $T\sim 2~{\rm MeV}$. However, the additional interaction between photons and neutrinos induced by a magnetic moment may lead to a delayed neutrino decoupling. This imposes an upper bound on $\mu_{\textrm{tr}}$~(see e.g.~\cite{Vassh:2015yza} for analogous active limits).

Our results for the NTMM scenario are shown in Fig.~\ref{fig:sens-mu}. The shaded regions are disfavored by past experiments as outlined above, while the contours correspond to the regions where more than one DB event would be expected at IceCube, for six years of data taking. The left panel shows the results for a NTMM between $N$ and $\nu_\tau$. Our results indicate that IceCube has the potential to improve more than two orders of magnitude over current constraints for NTMM, for $m_N \sim 1~\textrm{MeV} - 1~\textrm{GeV}$. The right panel, on the other hand, shows the results for a NTMM between $N$ and $\nu_\mu$. The computation of the number of events is identical as for $\nu_\tau-N$ transitions, replacing the oscillation probability $P_{\mu\tau}$ by $ P_{\mu\mu}$ in Eq.~\eqref{eq:events}. Even though current constraints are stronger in this case, we also find that IceCube could significantly improve over present bounds.

As a final remark, it is known that operators that generate large NTMM may also induce large corrections to neutrino masses, leading to a fine-tuning problem~\cite{Barbieri:1987xm}. Nevertheless, in simple scenarios like the Zee model~\cite{Zee:1980ai} it is possible to obtain $\mu_{\rm tr}\sim\mathcal{O}(10^{-10}\mu_B)$ between active neutrinos without any tuning~\cite{Babu:1992vq}. Moreover, 
the NTMM operator in Eq.~\eqref{eq:operator} will contribute to the Dirac $\nu_L-N_R$ mass, which is allowed to be much larger than $m_\nu$ in Seesaw scenarios, for instance.

%

%
\textit{Conclusions.}
In this letter, we have studied the potential of the IceCube detector to look for new physics using low-energy DB events. The collaboration has already performed searches for events with this topology at ultra-high energies, which are expected in the SM from the CC interactions of PeV tau neutrinos. In this work we have shown how very simple new physics scenarios with GeV-scale right-handed neutrinos would lead to a similar topology, with two low-energy cascades that could be spatially resolved in the detector. We find that IceCube may be able to improve by orders of magnitude the current constraints on the two scenarios considered here.

\textit{Acknowledgements.}
We warmly thank Tyce de Young for useful discussions on the IceCube detector performance. We are very grateful as well for insightful discussions with Kaladi Babu, Enrique Fernandez-Martinez, Jacobo Lopez-Pavon, Kohta Murase and Josef Pradler. This work received partial support from the European Union through the Elusives (H2020-MSCA-ITN-2015-674896) and InvisiblesPlus (H2020-MSCA-RISE-2015-690575) grants. I.M.S. is very grateful to the University of South Dakota for its support.  I.M-S. acknowledges support through the Spanish grants FPA2015-65929-P (MINECO/FEDER, UE) and the Spanish Research Agency (Agencia Estatal de Investigaci\'on) through the grants IFT ``Centro de Excelencia Severo Ochoa'' SEV-2012-0249 and SEV-2016-0597, and would like to thank the Fermilab theory department for their kind hospitality during his visits, where this work was started. This manuscript has been authored by Fermi Research Alliance, LLC under Contract No. DE-AC02-07CH11359 with the U.S. Department of Energy, Office of Science, Office of High Energy Physics. The publisher, by accepting the article for publication, acknowledges that the United States Government retains a non-exclusive, paid-up, irrevocable, world-wide license to publish or reproduce the published form of this manuscript, or allow others to do so, for United States Government purposes. 

\newpage
\section{Supplemental Material}

\subsection{Deep Inelastic Scattering Cross Section}

The fully inclusive cross section from a neutrino transition magnetic moment in the DIS regime can be written in terms of the leptonic $L_{\mu \nu}$ and hadronic $W^{\mu \nu}$ matrix elements

\begin{widetext}
\[
\begin{aligned}
&&L_{\mu \nu} = \frac{1}{8}{\rm Tr}\left[(\slashed{k} -m_{N}) [\slashed{q},\gamma^{\mu}] P_{L} \slashed{k} [\gamma^{\nu}, \slashed{q}] P_{R}\right] \\
W_{\mu \nu}&=& W_{1}(Q^{2}, \nu) \left(-g_{\mu \nu} + \frac{q_{\mu} q_{\nu}}{q^{2}}\right) +  \frac{W_{2}(Q^{2}, \nu)}{M^2} \left(p_{\mu} - \frac{(p \cdot q)}{q^{2}}q^{\mu}\right)\left(p_{\nu} - \frac{(p \cdot q)}{q^{2}}q^{\nu}\right),
\end{aligned}
\]
\end{widetext}
where $M$ is the  target mass and $m_{N}$ is the up-scattered heavy neutrino mass. Here $k$ ($k'$) are the incoming (outgoing) neutrino momenta while $p$ ($p'$) are the initial (final) state target momenta. 

With the usual DIS variables
\bea q &=& k- k', \\
Q^{2} &=& -q^{2} = 4 E_{\nu} E_{\nu}' \sin^{2} (\theta/2)~~~~({\rm lab~frame}),\\
\nu &=& (p \cdot q)/M = E_{\nu} - E_{\nu}'~~~~({\rm lab~frame}),\\
x&=& Q^{2}/(2M \nu), \\
y&=& \nu/E_{\nu},
\eea
it is found that 
\be
L_{\mu \nu}W^{\mu \nu} = 2ME_\nu^2 x \left[ 2xy^2 MW_1-y(y-2)^2E_\nu W_2\right].
\ee
The differential cross section in the laboratory frame is given by
\begin{equation}
 \frac{d^2\sigma}{dE_\nu'd_\Omega'}  = \frac{1}{16\pi^2}\frac{E_\nu'}{E_\nu}L_{\mu \nu}W^{\mu \nu}.
\end{equation}
Parametrizing the form factors $W_{1,2}$ by
\begin{equation}
  W_1(Q^2,\nu)=\frac{\sum_i e_i^2 f_i(x)}{2M},\,\,\, W_2(Q^2,\nu)=\frac{\sum_i e_i^2 x f_i(x)}{\nu}\nonumber
\end{equation}
allow us to write the differential cross section as 
\begin{equation}
  \frac{d^2\sigma}{dxdy}\simeq 16\pi\alpha \mu_{\rm tr}^2 \left(\frac{1-y}{y}\right) \sum_i e_i^2 f_i(x).
  \end{equation}

\bibliographystyle{apsrev4-1}

\bibliography{nu}

\begin{thebibliography}{35}%
\makeatletter
\providecommand \@ifxundefined [1]{%
 \@ifx{#1\undefined}
}%
\providecommand \@ifnum [1]{%
 \ifnum #1\expandafter \@firstoftwo
 \else \expandafter \@secondoftwo
 \fi
}%
\providecommand \@ifx [1]{%
 \ifx #1\expandafter \@firstoftwo
 \else \expandafter \@secondoftwo
 \fi
}%
\providecommand \natexlab [1]{#1}%
\providecommand \enquote  [1]{``#1''}%
\providecommand \bibnamefont  [1]{#1}%
\providecommand \bibfnamefont [1]{#1}%
\providecommand \citenamefont [1]{#1}%
\providecommand \href@noop [0]{\@secondoftwo}%
\providecommand \href [0]{\begingroup \@sanitize@url \@href}%
\providecommand \@href[1]{\@@startlink{#1}\@@href}%
\providecommand \@@href[1]{\endgroup#1\@@endlink}%
\providecommand \@sanitize@url [0]{\catcode `\\12\catcode `\$12\catcode
  `\&12\catcode `\#12\catcode `\^12\catcode `\_12\catcode `\%12\relax}%
\providecommand \@@startlink[1]{}%
\providecommand \@@endlink[0]{}%
\providecommand \url  [0]{\begingroup\@sanitize@url \@url }%
\providecommand \@url [1]{\endgroup\@href {#1}{\urlprefix }}%
\providecommand \urlprefix  [0]{URL }%
\providecommand \Eprint [0]{\href }%
\providecommand \doibase [0]{http://dx.doi.org/}%
\providecommand \selectlanguage [0]{\@gobble}%
\providecommand \bibinfo  [0]{\@secondoftwo}%
\providecommand \bibfield  [0]{\@secondoftwo}%
\providecommand \translation [1]{[#1]}%
\providecommand \BibitemOpen [0]{}%
\providecommand \bibitemStop [0]{}%
\providecommand \bibitemNoStop [0]{.\EOS\space}%
\providecommand \EOS [0]{\spacefactor3000\relax}%
\providecommand \BibitemShut  [1]{\csname bibitem#1\endcsname}%
\let\auto@bib@innerbib\@empty
\bibitem [{\citenamefont {Minkowski}(1977)}]{Minkowski:1977sc}%
  \BibitemOpen
  \bibfield  {author} {\bibinfo {author} {\bibfnamefont {P.}~\bibnamefont
  {Minkowski}},\ }\href {\doibase 10.1016/0370-2693(77)90435-X} {\bibfield
  {journal} {\bibinfo  {journal} {Phys. Lett.}\ }\textbf {\bibinfo {volume}
  {B67}},\ \bibinfo {pages} {421} (\bibinfo {year} {1977})}\BibitemShut
  {NoStop}%
\bibitem [{\citenamefont {Gell-Mann}\ \emph {et~al.}(1979)\citenamefont
  {Gell-Mann}, \citenamefont {Ramond},\ and\ \citenamefont
  {Slansky}}]{GellMann:1980vs}%
  \BibitemOpen
  \bibfield  {author} {\bibinfo {author} {\bibfnamefont {M.}~\bibnamefont
  {Gell-Mann}}, \bibinfo {author} {\bibfnamefont {P.}~\bibnamefont {Ramond}}, \
  and\ \bibinfo {author} {\bibfnamefont {R.}~\bibnamefont {Slansky}},\
  }\bibfield  {booktitle} {\emph {\bibinfo {booktitle} {{Supergravity Workshop
  Stony Brook, New York, September 27-28, 1979}}},\ }\href@noop {} {\bibfield
  {journal} {\bibinfo  {journal} {Conf. Proc.}\ }\textbf {\bibinfo {volume}
  {C790927}},\ \bibinfo {pages} {315} (\bibinfo {year} {1979})},\ \Eprint
  {http://arxiv.org/abs/1306.4669} {arXiv:1306.4669 [hep-th]} \BibitemShut
  {NoStop}%
\bibitem [{\citenamefont {Mohapatra}\ and\ \citenamefont
  {Senjanovic}(1980)}]{Mohapatra:1979ia}%
  \BibitemOpen
  \bibfield  {author} {\bibinfo {author} {\bibfnamefont {R.~N.}\ \bibnamefont
  {Mohapatra}}\ and\ \bibinfo {author} {\bibfnamefont {G.}~\bibnamefont
  {Senjanovic}},\ }\href {\doibase 10.1103/PhysRevLett.44.912} {\bibfield
  {journal} {\bibinfo  {journal} {Phys. Rev. Lett.}\ }\textbf {\bibinfo
  {volume} {44}},\ \bibinfo {pages} {912} (\bibinfo {year} {1980})}\BibitemShut
  {NoStop}%
\bibitem [{\citenamefont {Vissani}(1998)}]{Vissani:1997ys}%
  \BibitemOpen
  \bibfield  {author} {\bibinfo {author} {\bibfnamefont {F.}~\bibnamefont
  {Vissani}},\ }\href {\doibase 10.1103/PhysRevD.57.7027} {\bibfield  {journal}
  {\bibinfo  {journal} {Phys. Rev.}\ }\textbf {\bibinfo {volume} {D57}},\
  \bibinfo {pages} {7027} (\bibinfo {year} {1998})},\ \Eprint
  {http://arxiv.org/abs/hep-ph/9709409} {arXiv:hep-ph/9709409 [hep-ph]}
  \BibitemShut {NoStop}%
\bibitem [{\citenamefont {Kusenko}(2009)}]{Kusenko:2009up}%
  \BibitemOpen
  \bibfield  {author} {\bibinfo {author} {\bibfnamefont {A.}~\bibnamefont
  {Kusenko}},\ }\href {\doibase 10.1016/j.physrep.2009.07.004} {\bibfield
  {journal} {\bibinfo  {journal} {Phys. Rept.}\ }\textbf {\bibinfo {volume}
  {481}},\ \bibinfo {pages} {1} (\bibinfo {year} {2009})},\ \Eprint
  {http://arxiv.org/abs/0906.2968} {arXiv:0906.2968 [hep-ph]} \BibitemShut
  {NoStop}%
\bibitem [{\citenamefont {Asaka}\ \emph {et~al.}(2005)\citenamefont {Asaka},
  \citenamefont {Blanchet},\ and\ \citenamefont {Shaposhnikov}}]{Asaka:2005an}%
  \BibitemOpen
  \bibfield  {author} {\bibinfo {author} {\bibfnamefont {T.}~\bibnamefont
  {Asaka}}, \bibinfo {author} {\bibfnamefont {S.}~\bibnamefont {Blanchet}}, \
  and\ \bibinfo {author} {\bibfnamefont {M.}~\bibnamefont {Shaposhnikov}},\
  }\href {\doibase 10.1016/j.physletb.2005.09.070} {\bibfield  {journal}
  {\bibinfo  {journal} {Phys. Lett.}\ }\textbf {\bibinfo {volume} {B631}},\
  \bibinfo {pages} {151} (\bibinfo {year} {2005})},\ \Eprint
  {http://arxiv.org/abs/hep-ph/0503065} {arXiv:hep-ph/0503065 [hep-ph]}
  \BibitemShut {NoStop}%
\bibitem [{\citenamefont {Asaka}\ and\ \citenamefont
  {Shaposhnikov}(2005)}]{Asaka:2005pn}%
  \BibitemOpen
  \bibfield  {author} {\bibinfo {author} {\bibfnamefont {T.}~\bibnamefont
  {Asaka}}\ and\ \bibinfo {author} {\bibfnamefont {M.}~\bibnamefont
  {Shaposhnikov}},\ }\href {\doibase 10.1016/j.physletb.2005.06.020} {\bibfield
   {journal} {\bibinfo  {journal} {Phys. Lett.}\ }\textbf {\bibinfo {volume}
  {B620}},\ \bibinfo {pages} {17} (\bibinfo {year} {2005})},\ \Eprint
  {http://arxiv.org/abs/hep-ph/0505013} {arXiv:hep-ph/0505013 [hep-ph]}
  \BibitemShut {NoStop}%
\bibitem [{\citenamefont {Hern{\'a}ndez}\ \emph {et~al.}(2016)\citenamefont
  {Hern{\'a}ndez}, \citenamefont {Kekic}, \citenamefont {L{\'o}pez-Pav{\'o}n},
  \citenamefont {Racker},\ and\ \citenamefont {Salvado}}]{Hernandez:2016kel}%
  \BibitemOpen
  \bibfield  {author} {\bibinfo {author} {\bibfnamefont {P.}~\bibnamefont
  {Hern{\'a}ndez}}, \bibinfo {author} {\bibfnamefont {M.}~\bibnamefont
  {Kekic}}, \bibinfo {author} {\bibfnamefont {J.}~\bibnamefont
  {L{\'o}pez-Pav{\'o}n}}, \bibinfo {author} {\bibfnamefont {J.}~\bibnamefont
  {Racker}}, \ and\ \bibinfo {author} {\bibfnamefont {J.}~\bibnamefont
  {Salvado}},\ }\href {\doibase 10.1007/JHEP08(2016)157} {\bibfield  {journal}
  {\bibinfo  {journal} {JHEP}\ }\textbf {\bibinfo {volume} {08}},\ \bibinfo
  {pages} {157} (\bibinfo {year} {2016})},\ \Eprint
  {http://arxiv.org/abs/1606.06719} {arXiv:1606.06719 [hep-ph]} \BibitemShut
  {NoStop}%
\bibitem [{\citenamefont {Hern{\'a}ndez}\ \emph {et~al.}(2015)\citenamefont
  {Hern{\'a}ndez}, \citenamefont {Kekic}, \citenamefont {L{\'o}pez-Pav{\'o}n},
  \citenamefont {Racker},\ and\ \citenamefont {Rius}}]{Hernandez:2015wna}%
  \BibitemOpen
  \bibfield  {author} {\bibinfo {author} {\bibfnamefont {P.}~\bibnamefont
  {Hern{\'a}ndez}}, \bibinfo {author} {\bibfnamefont {M.}~\bibnamefont
  {Kekic}}, \bibinfo {author} {\bibfnamefont {J.}~\bibnamefont
  {L{\'o}pez-Pav{\'o}n}}, \bibinfo {author} {\bibfnamefont {J.}~\bibnamefont
  {Racker}}, \ and\ \bibinfo {author} {\bibfnamefont {N.}~\bibnamefont
  {Rius}},\ }\href {\doibase 10.1007/JHEP10(2015)067} {\bibfield  {journal}
  {\bibinfo  {journal} {JHEP}\ }\textbf {\bibinfo {volume} {10}},\ \bibinfo
  {pages} {067} (\bibinfo {year} {2015})},\ \Eprint
  {http://arxiv.org/abs/1508.03676} {arXiv:1508.03676 [hep-ph]} \BibitemShut
  {NoStop}%
\bibitem [{\citenamefont {Antusch}\ and\ \citenamefont
  {Fischer}(2014)}]{Antusch:2014woa}%
  \BibitemOpen
  \bibfield  {author} {\bibinfo {author} {\bibfnamefont {S.}~\bibnamefont
  {Antusch}}\ and\ \bibinfo {author} {\bibfnamefont {O.}~\bibnamefont
  {Fischer}},\ }\href {\doibase 10.1007/JHEP10(2014)094} {\bibfield  {journal}
  {\bibinfo  {journal} {JHEP}\ }\textbf {\bibinfo {volume} {10}},\ \bibinfo
  {pages} {094} (\bibinfo {year} {2014})},\ \Eprint
  {http://arxiv.org/abs/1407.6607} {arXiv:1407.6607 [hep-ph]} \BibitemShut
  {NoStop}%
\bibitem [{\citenamefont {Fernandez-Martinez}\ \emph
  {et~al.}(2016)\citenamefont {Fernandez-Martinez}, \citenamefont
  {Hernandez-Garcia},\ and\ \citenamefont
  {Lopez-Pavon}}]{Fernandez-Martinez:2016lgt}%
  \BibitemOpen
  \bibfield  {author} {\bibinfo {author} {\bibfnamefont {E.}~\bibnamefont
  {Fernandez-Martinez}}, \bibinfo {author} {\bibfnamefont {J.}~\bibnamefont
  {Hernandez-Garcia}}, \ and\ \bibinfo {author} {\bibfnamefont
  {J.}~\bibnamefont {Lopez-Pavon}},\ }\href {\doibase 10.1007/JHEP08(2016)033}
  {\bibfield  {journal} {\bibinfo  {journal} {JHEP}\ }\textbf {\bibinfo
  {volume} {08}},\ \bibinfo {pages} {033} (\bibinfo {year} {2016})},\ \Eprint
  {http://arxiv.org/abs/1605.08774} {arXiv:1605.08774 [hep-ph]} \BibitemShut
  {NoStop}%
\bibitem [{\citenamefont {Atre}\ \emph {et~al.}(2009)\citenamefont {Atre},
  \citenamefont {Han}, \citenamefont {Pascoli},\ and\ \citenamefont
  {Zhang}}]{Atre:2009rg}%
  \BibitemOpen
  \bibfield  {author} {\bibinfo {author} {\bibfnamefont {A.}~\bibnamefont
  {Atre}}, \bibinfo {author} {\bibfnamefont {T.}~\bibnamefont {Han}}, \bibinfo
  {author} {\bibfnamefont {S.}~\bibnamefont {Pascoli}}, \ and\ \bibinfo
  {author} {\bibfnamefont {B.}~\bibnamefont {Zhang}},\ }\href {\doibase
  10.1088/1126-6708/2009/05/030} {\bibfield  {journal} {\bibinfo  {journal}
  {JHEP}\ }\textbf {\bibinfo {volume} {05}},\ \bibinfo {pages} {030} (\bibinfo
  {year} {2009})},\ \Eprint {http://arxiv.org/abs/0901.3589} {arXiv:0901.3589
  [hep-ph]} \BibitemShut {NoStop}%
\bibitem [{\citenamefont {Learned}\ and\ \citenamefont
  {Pakvasa}(1995)}]{Learned:1994wg}%
  \BibitemOpen
  \bibfield  {author} {\bibinfo {author} {\bibfnamefont {J.~G.}\ \bibnamefont
  {Learned}}\ and\ \bibinfo {author} {\bibfnamefont {S.}~\bibnamefont
  {Pakvasa}},\ }\href {\doibase 10.1016/0927-6505(94)00043-3} {\bibfield
  {journal} {\bibinfo  {journal} {Astropart. Phys.}\ }\textbf {\bibinfo
  {volume} {3}},\ \bibinfo {pages} {267} (\bibinfo {year} {1995})},\ \Eprint
  {http://arxiv.org/abs/hep-ph/9405296} {arXiv:hep-ph/9405296 [hep-ph]}
  \BibitemShut {NoStop}%
\bibitem [{\citenamefont {Aartsen}\ \emph {et~al.}(2015)\citenamefont {Aartsen}
  \emph {et~al.}}]{Aartsen:2015dlt}%
  \BibitemOpen
  \bibfield  {author} {\bibinfo {author} {\bibfnamefont {M.~G.}\ \bibnamefont
  {Aartsen}} \emph {et~al.} (\bibinfo {collaboration} {IceCube}),\ }\href@noop
  {} {\  (\bibinfo {year} {2015})},\ \Eprint {http://arxiv.org/abs/1509.06212}
  {arXiv:1509.06212 [astro-ph.HE]} \BibitemShut {NoStop}%
\bibitem [{\citenamefont {Abreu}\ \emph {et~al.}(1997)\citenamefont {Abreu}
  \emph {et~al.}}]{Abreu:1996pa}%
  \BibitemOpen
  \bibfield  {author} {\bibinfo {author} {\bibfnamefont {P.}~\bibnamefont
  {Abreu}} \emph {et~al.} (\bibinfo {collaboration} {DELPHI}),\ }\href
  {\doibase 10.1007/s002880050370} {\bibfield  {journal} {\bibinfo  {journal}
  {Z. Phys.}\ }\textbf {\bibinfo {volume} {C74}},\ \bibinfo {pages} {57}
  (\bibinfo {year} {1997})},\ \bibinfo {note} {[Erratum: Z.
  Phys.C75,580(1997)]}\BibitemShut {NoStop}%
\bibitem [{\citenamefont {Orloff}\ \emph {et~al.}(2002)\citenamefont {Orloff},
  \citenamefont {Rozanov},\ and\ \citenamefont {Santoni}}]{Orloff:2002de}%
  \BibitemOpen
  \bibfield  {author} {\bibinfo {author} {\bibfnamefont {J.}~\bibnamefont
  {Orloff}}, \bibinfo {author} {\bibfnamefont {A.~N.}\ \bibnamefont {Rozanov}},
  \ and\ \bibinfo {author} {\bibfnamefont {C.}~\bibnamefont {Santoni}},\ }\href
  {\doibase 10.1016/S0370-2693(02)02769-7} {\bibfield  {journal} {\bibinfo
  {journal} {Phys. Lett.}\ }\textbf {\bibinfo {volume} {B550}},\ \bibinfo
  {pages} {8} (\bibinfo {year} {2002})},\ \Eprint
  {http://arxiv.org/abs/hep-ph/0208075} {arXiv:hep-ph/0208075 [hep-ph]}
  \BibitemShut {NoStop}%
\bibitem [{\citenamefont {Aartsen}\ \emph
  {et~al.}(2017{\natexlab{a}})\citenamefont {Aartsen} \emph
  {et~al.}}]{Aartsen:2017nmd}%
  \BibitemOpen
  \bibfield  {author} {\bibinfo {author} {\bibfnamefont {M.~G.}\ \bibnamefont
  {Aartsen}} \emph {et~al.} (\bibinfo {collaboration} {IceCube}),\ }\href@noop
  {} {\  (\bibinfo {year} {2017}{\natexlab{a}})},\ \Eprint
  {http://arxiv.org/abs/1707.07081} {arXiv:1707.07081 [hep-ex]} \BibitemShut
  {NoStop}%
\bibitem [{\citenamefont {Gorbunov}\ and\ \citenamefont
  {Shaposhnikov}(2007)}]{Gorbunov:2007ak}%
  \BibitemOpen
  \bibfield  {author} {\bibinfo {author} {\bibfnamefont {D.}~\bibnamefont
  {Gorbunov}}\ and\ \bibinfo {author} {\bibfnamefont {M.}~\bibnamefont
  {Shaposhnikov}},\ }\href {\doibase 10.1007/JHEP11(2013)101,
  10.1088/1126-6708/2007/10/015} {\bibfield  {journal} {\bibinfo  {journal}
  {JHEP}\ }\textbf {\bibinfo {volume} {10}},\ \bibinfo {pages} {015} (\bibinfo
  {year} {2007})},\ \bibinfo {note} {[Erratum: JHEP11,101(2013)]},\ \Eprint
  {http://arxiv.org/abs/0705.1729} {arXiv:0705.1729 [hep-ph]} \BibitemShut
  {NoStop}%
\bibitem [{\citenamefont {Honda}\ \emph {et~al.}(2015)\citenamefont {Honda},
  \citenamefont {Sajjad~Athar}, \citenamefont {Kajita}, \citenamefont
  {Kasahara},\ and\ \citenamefont {Midorikawa}}]{Honda:2015fha}%
  \BibitemOpen
  \bibfield  {author} {\bibinfo {author} {\bibfnamefont {M.}~\bibnamefont
  {Honda}}, \bibinfo {author} {\bibfnamefont {M.}~\bibnamefont {Sajjad~Athar}},
  \bibinfo {author} {\bibfnamefont {T.}~\bibnamefont {Kajita}}, \bibinfo
  {author} {\bibfnamefont {K.}~\bibnamefont {Kasahara}}, \ and\ \bibinfo
  {author} {\bibfnamefont {S.}~\bibnamefont {Midorikawa}},\ }\href {\doibase
  10.1103/PhysRevD.92.023004} {\bibfield  {journal} {\bibinfo  {journal} {Phys.
  Rev.}\ }\textbf {\bibinfo {volume} {D92}},\ \bibinfo {pages} {023004}
  (\bibinfo {year} {2015})},\ \Eprint {http://arxiv.org/abs/1502.03916}
  {arXiv:1502.03916 [astro-ph.HE]} \BibitemShut {NoStop}%
\bibitem [{\citenamefont {Aartsen}\ \emph
  {et~al.}(2017{\natexlab{b}})\citenamefont {Aartsen} \emph
  {et~al.}}]{Aartsen:2016nxy}%
  \BibitemOpen
  \bibfield  {author} {\bibinfo {author} {\bibfnamefont {M.~G.}\ \bibnamefont
  {Aartsen}} \emph {et~al.} (\bibinfo {collaboration} {IceCube}),\ }\href
  {\doibase 10.1088/1748-0221/12/03/P03012} {\bibfield  {journal} {\bibinfo
  {journal} {JINST}\ }\textbf {\bibinfo {volume} {12}},\ \bibinfo {pages}
  {P03012} (\bibinfo {year} {2017}{\natexlab{b}})},\ \Eprint
  {http://arxiv.org/abs/1612.05093} {arXiv:1612.05093 [astro-ph.IM]}
  \BibitemShut {NoStop}%
\bibitem [{\citenamefont {Ge}\ \emph {et~al.}(2017)\citenamefont {Ge},
  \citenamefont {Lindner},\ and\ \citenamefont {Rodejohann}}]{Ge:2017poy}%
  \BibitemOpen
  \bibfield  {author} {\bibinfo {author} {\bibfnamefont {S.-F.}\ \bibnamefont
  {Ge}}, \bibinfo {author} {\bibfnamefont {M.}~\bibnamefont {Lindner}}, \ and\
  \bibinfo {author} {\bibfnamefont {W.}~\bibnamefont {Rodejohann}},\
  }\href@noop {} {\  (\bibinfo {year} {2017})},\ \Eprint
  {http://arxiv.org/abs/1702.02617} {arXiv:1702.02617 [hep-ph]} \BibitemShut
  {NoStop}%
\bibitem [{\citenamefont {Hignight}(2017)}]{LakeLouiseTalk}%
  \BibitemOpen
  \bibfield  {author} {\bibinfo {author} {\bibfnamefont {J.}~\bibnamefont
  {Hignight}},\ }\href@noop {} {\emph {\bibinfo {title} {Measurements of
  atmospheric NuMu disappearance with IceCube-DeepCore}}} (\bibinfo {year}
  {2017}),\ \bibinfo {note} {talk given at the Lake Louise Winter Institute
  2017,
  \url{https://indico.cern.ch/event/531113/contributions/2430431/}}\BibitemShut
  {NoStop}%
\bibitem [{\citenamefont {Cveti{\v c}}\ \emph {et~al.}(2017)\citenamefont
  {Cveti{\v c}}, \citenamefont {Halzen}, \citenamefont {Kim},\ and\
  \citenamefont {Oh}}]{Cvetic:2017gkt}%
  \BibitemOpen
  \bibfield  {author} {\bibinfo {author} {\bibfnamefont {G.}~\bibnamefont
  {Cveti{\v c}}}, \bibinfo {author} {\bibfnamefont {F.}~\bibnamefont {Halzen}},
  \bibinfo {author} {\bibfnamefont {C.~S.}\ \bibnamefont {Kim}}, \ and\
  \bibinfo {author} {\bibfnamefont {S.}~\bibnamefont {Oh}},\ }\href@noop {} {\
  (\bibinfo {year} {2017})},\ \Eprint {http://arxiv.org/abs/1702.04335}
  {arXiv:1702.04335 [hep-ph]} \BibitemShut {NoStop}%
\bibitem [{\citenamefont {Schwienhorst}\ \emph {et~al.}(2001)\citenamefont
  {Schwienhorst} \emph {et~al.}}]{Schwienhorst:2001sj}%
  \BibitemOpen
  \bibfield  {author} {\bibinfo {author} {\bibfnamefont {R.}~\bibnamefont
  {Schwienhorst}} \emph {et~al.} (\bibinfo {collaboration} {DONUT}),\ }\href
  {\doibase 10.1016/S0370-2693(01)00746-8} {\bibfield  {journal} {\bibinfo
  {journal} {Phys. Lett.}\ }\textbf {\bibinfo {volume} {B513}},\ \bibinfo
  {pages} {23} (\bibinfo {year} {2001})},\ \Eprint
  {http://arxiv.org/abs/hep-ex/0102026} {arXiv:hep-ex/0102026 [hep-ex]}
  \BibitemShut {NoStop}%
\bibitem [{\citenamefont {Zeller}\ \emph {et~al.}(2002)\citenamefont {Zeller}
  \emph {et~al.}}]{Zeller:2001hh}%
  \BibitemOpen
  \bibfield  {author} {\bibinfo {author} {\bibfnamefont {G.~P.}\ \bibnamefont
  {Zeller}} \emph {et~al.} (\bibinfo {collaboration} {NuTeV}),\ }\href
  {\doibase 10.1103/PhysRevLett.88.091802} {\bibfield  {journal} {\bibinfo
  {journal} {Phys. Rev. Lett.}\ }\textbf {\bibinfo {volume} {88}},\ \bibinfo
  {pages} {091802} (\bibinfo {year} {2002})},\ \bibinfo {note} {[Erratum: Phys.
  Rev. Lett.90,239902(2003)]},\ \Eprint {http://arxiv.org/abs/hep-ex/0110059}
  {arXiv:hep-ex/0110059 [hep-ex]} \BibitemShut {NoStop}%
\bibitem [{\citenamefont {Altegoer}\ \emph {et~al.}(1998)\citenamefont
  {Altegoer} \emph {et~al.}}]{Altegoer:1997gv}%
  \BibitemOpen
  \bibfield  {author} {\bibinfo {author} {\bibfnamefont {J.}~\bibnamefont
  {Altegoer}} \emph {et~al.} (\bibinfo {collaboration} {NOMAD}),\ }\href
  {\doibase 10.1016/S0168-9002(97)01079-6} {\bibfield  {journal} {\bibinfo
  {journal} {Nucl. Instrum. Meth.}\ }\textbf {\bibinfo {volume} {A404}},\
  \bibinfo {pages} {96} (\bibinfo {year} {1998})}\BibitemShut {NoStop}%
\bibitem [{\citenamefont {Gninenko}\ and\ \citenamefont
  {Krasnikov}(1999)}]{Gninenko:1998nn}%
  \BibitemOpen
  \bibfield  {author} {\bibinfo {author} {\bibfnamefont {S.~N.}\ \bibnamefont
  {Gninenko}}\ and\ \bibinfo {author} {\bibfnamefont {N.~V.}\ \bibnamefont
  {Krasnikov}},\ }\href {\doibase 10.1016/S0370-2693(99)00130-6} {\bibfield
  {journal} {\bibinfo  {journal} {Phys. Lett.}\ }\textbf {\bibinfo {volume}
  {B450}},\ \bibinfo {pages} {165} (\bibinfo {year} {1999})},\ \Eprint
  {http://arxiv.org/abs/hep-ph/9808370} {arXiv:hep-ph/9808370 [hep-ph]}
  \BibitemShut {NoStop}%
\bibitem [{\citenamefont {Bellini}\ \emph {et~al.}(2014)\citenamefont {Bellini}
  \emph {et~al.}}]{Bellini:2014uqa}%
  \BibitemOpen
  \bibfield  {author} {\bibinfo {author} {\bibfnamefont {G.}~\bibnamefont
  {Bellini}} \emph {et~al.} (\bibinfo {collaboration} {BOREXINO}),\ }\href
  {\doibase 10.1038/nature13702} {\bibfield  {journal} {\bibinfo  {journal}
  {Nature}\ }\textbf {\bibinfo {volume} {512}},\ \bibinfo {pages} {383}
  (\bibinfo {year} {2014})}\BibitemShut {NoStop}%
\bibitem [{\citenamefont {Geiregat}\ \emph {et~al.}(1989)\citenamefont
  {Geiregat} \emph {et~al.}}]{Geiregat:1989sz}%
  \BibitemOpen
  \bibfield  {author} {\bibinfo {author} {\bibfnamefont {D.}~\bibnamefont
  {Geiregat}} \emph {et~al.} (\bibinfo {collaboration} {CHARM-II}),\ }\href
  {\doibase 10.1016/0370-2693(89)90457-7} {\bibfield  {journal} {\bibinfo
  {journal} {Phys. Lett.}\ }\textbf {\bibinfo {volume} {B232}},\ \bibinfo
  {pages} {539} (\bibinfo {year} {1989})}\BibitemShut {NoStop}%
\bibitem [{\citenamefont {Decamp}\ \emph {et~al.}(1992)\citenamefont {Decamp}
  \emph {et~al.}}]{Decamp:1991uy}%
  \BibitemOpen
  \bibfield  {author} {\bibinfo {author} {\bibfnamefont {D.}~\bibnamefont
  {Decamp}} \emph {et~al.} (\bibinfo {collaboration} {ALEPH}),\ }\href
  {\doibase 10.1016/0370-1573(92)90177-2} {\bibfield  {journal} {\bibinfo
  {journal} {Phys. Rept.}\ }\textbf {\bibinfo {volume} {216}},\ \bibinfo
  {pages} {253} (\bibinfo {year} {1992})}\BibitemShut {NoStop}%
\bibitem [{\citenamefont {Gninenko}(2011)}]{Gninenko:2010pr}%
  \BibitemOpen
  \bibfield  {author} {\bibinfo {author} {\bibfnamefont {S.~N.}\ \bibnamefont
  {Gninenko}},\ }\href {\doibase 10.1103/PhysRevD.83.015015} {\bibfield
  {journal} {\bibinfo  {journal} {Phys. Rev.}\ }\textbf {\bibinfo {volume}
  {D83}},\ \bibinfo {pages} {015015} (\bibinfo {year} {2011})},\ \Eprint
  {http://arxiv.org/abs/1009.5536} {arXiv:1009.5536 [hep-ph]} \BibitemShut
  {NoStop}%
\bibitem [{\citenamefont {Vassh}\ \emph {et~al.}(2015)\citenamefont {Vassh},
  \citenamefont {Grohs}, \citenamefont {Balantekin},\ and\ \citenamefont
  {Fuller}}]{Vassh:2015yza}%
  \BibitemOpen
  \bibfield  {author} {\bibinfo {author} {\bibfnamefont {N.}~\bibnamefont
  {Vassh}}, \bibinfo {author} {\bibfnamefont {E.}~\bibnamefont {Grohs}},
  \bibinfo {author} {\bibfnamefont {A.~B.}\ \bibnamefont {Balantekin}}, \ and\
  \bibinfo {author} {\bibfnamefont {G.~M.}\ \bibnamefont {Fuller}},\ }\href
  {\doibase 10.1103/PhysRevD.92.125020} {\bibfield  {journal} {\bibinfo
  {journal} {Phys. Rev.}\ }\textbf {\bibinfo {volume} {D92}},\ \bibinfo {pages}
  {125020} (\bibinfo {year} {2015})},\ \Eprint
  {http://arxiv.org/abs/1510.00428} {arXiv:1510.00428 [astro-ph.CO]}
  \BibitemShut {NoStop}%
\bibitem [{\citenamefont {Barbieri}\ and\ \citenamefont
  {Fiorentini}(1988)}]{Barbieri:1987xm}%
  \BibitemOpen
  \bibfield  {author} {\bibinfo {author} {\bibfnamefont {R.}~\bibnamefont
  {Barbieri}}\ and\ \bibinfo {author} {\bibfnamefont {G.}~\bibnamefont
  {Fiorentini}},\ }\href {\doibase 10.1016/0550-3213(88)90661-X} {\bibfield
  {journal} {\bibinfo  {journal} {Nucl. Phys.}\ }\textbf {\bibinfo {volume}
  {B304}},\ \bibinfo {pages} {909} (\bibinfo {year} {1988})}\BibitemShut
  {NoStop}%
\bibitem [{\citenamefont {Zee}(1980)}]{Zee:1980ai}%
  \BibitemOpen
  \bibfield  {author} {\bibinfo {author} {\bibfnamefont {A.}~\bibnamefont
  {Zee}},\ }\href {\doibase 10.1016/0370-2693(80)90349-4,
  10.1016/0370-2693(80)90193-8} {\bibfield  {journal} {\bibinfo  {journal}
  {Phys. Lett.}\ }\textbf {\bibinfo {volume} {93B}},\ \bibinfo {pages} {389}
  (\bibinfo {year} {1980})},\ \bibinfo {note} {[Erratum: Phys.
  Lett.95B,461(1980)]}\BibitemShut {NoStop}%
\bibitem [{\citenamefont {Babu}\ \emph {et~al.}(1992)\citenamefont {Babu},
  \citenamefont {Chang}, \citenamefont {Keung},\ and\ \citenamefont
  {Phillips}}]{Babu:1992vq}%
  \BibitemOpen
  \bibfield  {author} {\bibinfo {author} {\bibfnamefont {K.~S.}\ \bibnamefont
  {Babu}}, \bibinfo {author} {\bibfnamefont {D.}~\bibnamefont {Chang}},
  \bibinfo {author} {\bibfnamefont {W.-Y.}\ \bibnamefont {Keung}}, \ and\
  \bibinfo {author} {\bibfnamefont {I.}~\bibnamefont {Phillips}},\ }\href
  {\doibase 10.1103/PhysRevD.46.2268} {\bibfield  {journal} {\bibinfo
  {journal} {Phys. Rev.}\ }\textbf {\bibinfo {volume} {D46}},\ \bibinfo {pages}
  {2268} (\bibinfo {year} {1992})}\BibitemShut {NoStop}%
\end{thebibliography}%

\end{document}